\documentclass[11pt]{article}
\usepackage{graphicx,epsfig}
\usepackage{amsmath}
\usepackage{amsfonts}
\usepackage{amssymb}
\usepackage{units}
\usepackage{epsf}
\usepackage[square,numbers]{natbib}
\usepackage[margin=5pt,labelfont=bf,labelsep=colon]{caption}
\usepackage{graphicx}
\newcommand{\bm}[1]{\mbox{\boldmath$#1$}}
\newcommand{\be}{\begin{equation}}
\newcommand{\ee}{\end{equation}}
\newcommand{\ba}{\begin{align}}
\newcommand{\ea}{\end{align}}

\newcommand{\ket}[1]{\ensuremath{| \, #1 \, \rangle }}

\begin{document}
\title{Photon structure function revisited}
\author{Ch. Berger \thanks{\noindent email: berger@rwth-aachen.de}
\\{\small I. Physikalisches Institut der RWTH Aachen University, Germany}\date{}}

\maketitle

\begin{abstract}
The flux of papers from electron positron colliders containing data on the photon structure function  
$F_2^\gamma(x,Q^2)$ ended naturally around 2005. It is thus timely to review the theoretical basis
and confront the predictions with a summary of the experimental results. The discussion will focus on
the increase of the structure function 
with $x$ (for $x$ away from the boundaries) and its rise  with $\ln Q^2$,
both characteristics being dramatically different from hadronic structure functions.
The agreement of the experimental observations with the theoretical calculations  is a striking success of QCD. It also allows a new determination of the QCD coupling constant
$\alpha_S$  which very well corresponds to the values quoted in the literature. 
\end{abstract}

\section{Historical introduction}
The notion that hadron production in inelastic electron photon scattering can be described in 
terms of structure functions like in electron nucleon scattering is on first sight surprising because
photons are pointlike particles whereas nucleons have a radius of roughly $1$ fm. Nevertheless the 
concept makes sense, not because the photon consists of pions, quarks, gluons etc, but because it couples 
to other particles and thus can fluctuate e.g.\ into a quark antiquark pair or a $\rho$ meson.
These two basic processes are distinguished by the terms \emph{pointlike}
and \emph{hadronic} throughout the paper. The idea of a photon fluctuating into a $\rho$ meson or
other vector mesons  was soon applied  to estimate the inelastic $e\gamma$ scattering cross section  
in the vector meson dominance model~\cite{Walsh,BKT}. Calculating the structure function in the 
quark model~\cite{WZ}
then opened the intriguing possibility to investigate experimentally  a structure function rising towards
large $x$ and showing a distinctive scale breaking because of the proportionality to $\ln Q^2$.

Excitement rose after the first calculation of the leading order QCD corrections, 
because Witten~\cite{Witt} not only calculated the markedly different $x$ dependence
of the structure function in QCD 
but demonstrated that the QCD parameter $\Lambda$ could in principle be
determined by measuring an absolute cross section quite in contrast to lepton nucleon
scattering, where small scale breaking effects in the $Q^2$ evolution of the 
structure function have to be studied. This ``remarkable result''~\cite{Lsmith}  initiated
intensive discussions between theorists and experimentalists  and
passed the first experimental test~\cite{Pluto1} with flying colors. QCD calculations at
next to leading order~\cite{BB,Duke} allowed to give $\Lambda$ a precise meaning
in the $\overline{\rm MS}$ renormalization scheme, but also revealed a sickness of the absolute perturbative calculation, producing negative values of $F_2^\gamma$ near $x=0$.

In an invited  talk at the 1983 Aachen conference on photon photon collisions~\cite{DuAAc} 
the audience was warned that  
the implications of these discoveries for the experimental goal of a direct determination of
$\Lambda$ from $F_2^\gamma$ were not altogether positive 
despite ``the almost incredible advances on the experimental side''.
Instead it was recommended to utilize the $Q^2$ evolution like in deep inelastic scattering,
a program which was also pursued by other groups~\cite{GR2}. 

Ten years later 
an algebraic error in the original calculation~\cite{BB} was discovered.
Correcting this error~\cite{GRV,FoPi}
squeezed the negative spike near $x=0$ to very small $x$ values where it is of 
\hbox{negligible} practical importance.
For the same reason ad hoc attempts~\cite{AnGr} to cure the problem (although still in principle important)
proved to be  unnecessary for experimental analyses at NLO accuracy. 

A new approach to follow the original goal~\cite{AKS} showed promising results. However, based on the
results of~\cite{MVV,VMV} the structure function for virtual photons was calculated~\cite{USU} in next to next to leading order (NNLO). The findings of this investigation forces one to the conclusion that an absolute prediction for the structure function of real photons is unstable at the three loop level. The concern of the 1980's is thus still valid, albeit at a higher order in the perturbative series. 

\section{Basics}
\begin{figure}
\begin{center}
\includegraphics[width=0.55\textwidth]{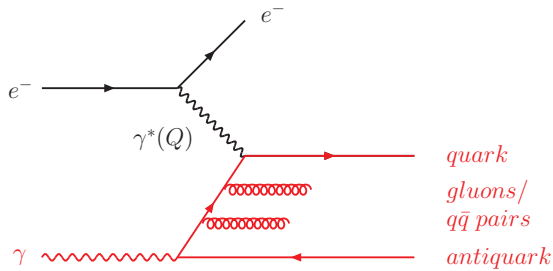}
\end{center}
\caption{ Electron-photon scattering [generic Feynman diagram]. The incoming target photon  
$\gamma$ splits into a nearly collinear quark-antiquark pair. In QCD, the momentum of the internal quark line is reduced by gluon radiation. The impinging electron is scattered off the quark to large angles, the scatter pattern revealing the internal quark structure of the photon. Quark, antiquark and gluons finally fragment to hadrons.}
\label{fig:Feyn}

\end{figure}

Deep-inelastic electron-photon scattering at high energies
\begin{equation}
   e^- + \gamma \to e^- + {\rm hadrons} \,,
\end{equation}
is characterized by a large momentum transfer $Q$ of the scattered electron and a large invariant mass $W$ of the hadrons. 
The electron energies $E_1$ and $E_1'$ in initial and final state combined with the scattering angle $\theta_1$ 
define the (negative) momentum transfer squared, $-Q^2$, on the electron line with
\begin{equation}    
   Q^2 = 4 E_1 E_1' \sin^2\theta_1/2   \,.
\end{equation}

$Q^2$ and the Bjorken scaling variable $x$  defined as
\be
x=\frac{Q^2}{Q^2+W^2}\enspace ,
\label{xydef}\ee
are the essential variables  for discussing  the dynamics of the scattering process as can 
be seen from 
%
%
the cross section formula corresponding to Fig.~{\ref{fig:Feyn}}
\be
\frac{d^2\sigma}{dQ^2dx}=\frac{2\pi\alpha^2}{xQ^4}\big([1+(1-y)^2]
F_2^\gamma-y^2F_L^\gamma\big)
\label{dsigdq2dx}\ee
which depends on the two structure functions $F_2^\gamma(x,Q^2)$ and 
$F_L^\gamma(x,Q^2)$. Here $F_2^\gamma$ is a linear combination 
$F_2^\gamma = 2x\, F_T^\gamma + F_L^\gamma$ of  $F_T^\gamma(x,Q^2)$, 
describing the exchange of transversely polarized virtual photons, and $F_L^\gamma(x,Q^2)$, 
associated with the exchange of longitudinally polarized virtual photons. 
The scaling variable $y$ used in the last equation is given by
\be
y=\frac{Q^2}{xs}
\ee
(with  $s=4EE^\gamma$) and 
can also be directly calculated  from $E_1,E'_1$ and $\theta_1$ via
\be
y =1 - E_1' / E_1 \cos^2\theta_1/2 \enspace .    
\ee 

For QED processes $e\gamma\rightarrow e\mu^+\mu^-$ in a region of phase space where
one of the muons travels along the direction of the incoming photon and the other is 
scattered at large angels (balancing the transverse momentum of the outgoing electron)
the structure function $F_2^\gamma$ has been calculated~\cite{Budnev,GRV} to be
\be
F_{2,{\rm QED}}^\gamma(x,Q^2)=\frac{\alpha}{\pi}x\left(h_1(x,Q^2)\ln{\frac{1+\beta}{1-\beta}}+h_2(x,Q^2)\right)
\label{F2QED}
\ee
with $\beta^2=1-4M^2x/(1-x)Q^2$ and $M$ denoting the muon mass.
The functions $h_1(x,Q^2)$ and $h_2(x,Q^2)$ are given by
\begin{eqnarray}
h_1&=&x^2+(1-x)^2+x(1-3x)\frac{4M^2}{Q^2}-x^2\frac{8M^4}{Q^4}\label{h1}\\
h_2&=&\beta\left(8x(1-x)-1-x(1-x)\frac{4M^2}{Q^2}\right)\label{h2}\enspace . 
\end{eqnarray}
For heavy quarks with three colors, fractional charge $e_q$ and masses
$M\gg \Lambda$ the right hand side of eq.~(\ref{F2QED}) has to be multiplied by $3e_q^4$.
Light quarks have current  masses with  $M\ll \Lambda$ (or at least $M< \Lambda$).
Neglecting all terms $\sim M^2/Q^2$  in eq.~(\ref{h1},\ref{h2}) and respecting
$\beta^2\rightarrow 1$ leads for each light quark to the expression
\be
F_2^\gamma(x,Q^2)=\frac{3\alpha}{\pi}e_q^4x\left( [x^2+(1-x)^2]
\ln{\frac{Q^2(1-x)}{xm_0^2}}
+8x(1-x)-1\right)
\label{qmv2}
\ee
where $m_0\approx 0.3$ GeV is a mass parameter somehow describing the confinement
of the light quarks~\cite{WZ}. The QCD parameter $\Lambda$ can be interpreted as an inverse
confinement radius. We therefore replace $m_0$ by $\Lambda$  and  keep 
in the leading log approximation only terms proportional to $\ln{Q^2}$  resulting
in
\be
F_2^\gamma(x,Q^2)=\frac{3\alpha}{\pi}\sum_{q=u,d,s}{e_q^4}x[x^2+(1-x)^2]
\ln{\frac{Q^2}{\Lambda^2}}
\label{qmv1}\ee
as the quark model or zero order QCD expression\footnote{A solution which formally requires very high
$Q^2$ (asymptotic solution) and/or staying away from the boundaries $x=0$ and $x=1$.}
for the photon structure function if only light quarks are
considered.

Using the general quark model relation
\be
F_2^\gamma(x,Q^2)=2x\sum_q e_q^2q^\gamma(x,Q^2)
\label{F2andq}\ee
connecting structure function and quark densities $q^\gamma$  one obtains for light quarks
the expression
\be
q^\gamma(x,Q^2)=\frac{3\alpha}{2\pi}e_q^2h_{\rm QM}(x)
\ln{\frac{Q^2}{\Lambda^2}}
\ee
with
\be
h_{\rm QM}(x)=x^2+(1-x)^2\enspace .
\label{hQM}\ee
The factor $2$ in eq.~(\ref{F2andq}) accounts for the fact that the photon contains
quarks and antiquarks with equal densities but the sum runs over quarks only.

\section{QCD predictions}
\subsection{Introduction, leading order calculations}

The first QCD analysis of the photon structure function~\cite{Witt} based on the
operator product expansion (OPE) gave a unified picture
of the hadronic and pointlike pieces including gluon radiation in leading order.
It was shown that the $x$ and $Q^2$ dependence of
$F_2^\gamma$ is unambiguously calculable for asymptotically high values
of $Q^2$. Due to the $1/\alpha_S$ term in front of the pointlike piece this in turn allows for
a determination of the strong coupling constant by measuring an absolute cross section. 
This unique result was later confirmed by calculations using a diagrammatic ansatz~\cite{Lsmith}
and/or solving the Altarelli-Parisi equations~\cite{DeWitt, Frazer} like in deep inelastic lepton nucleon 
scattering (DIS). A modern comprehensive summary of the theoretical foundations has been given by
Buras~\cite{Buras}.
In this paper we refer to~\cite{GRV}, where the $Q^2$ evolution equations for massless quarks 
(and gluons) are solved in the 
Mellin n-moment space in leading logarithmic order (LO) and next to leading logarithmic
order (NLO).

The $n$th Mellin moment of  a function $f(x,Q^2)$ is defined as
\be
f^n(Q^2)=\int_0^1x^{n-1}f(x,Q^2)dx\enspace .
\ee
For example, the quark model function $h_{\rm QM}(x)$ in eq.~(\ref{hQM}) is projected to
\be
h^n_{\rm QM}=\frac{n^2+n+2}{n(n+1)(n+2)}\enspace .
\ee
Like in DIS the quark densities are grouped into two classes
described by different evolution equations,
flavor non-singlet ({\rm NS}) and singlet ($\Sigma$):
\begin{eqnarray}
q^\gamma_{\rm NS}&=&\sum_q(e_q^2-\langle 
e^2\rangle)(q^\gamma+\bar{q}^\gamma)\nonumber\\
\Sigma^\gamma&=&\sum_q(q^\gamma+\bar{q}^\gamma)\enspace ,
\end{eqnarray}
where $\Sigma^\gamma$ is the first element of a two component singlet parton density
$q^\gamma_{\rm S}$, composed of  $\Sigma^\gamma$ and the gluon density $G^\gamma$.
Here $\langle e^2\rangle$ is the average charge squared in a  system with $f$ quark flavors,
e.g.\,$\langle e^2\rangle=2/9$ for $f=3$. 
The connection between structure function and quark densities is in LO defined by
\be
F_2^\gamma(x,Q^2)=x\left(q^\gamma_{\rm NS}(x,Q^2)+\langle e^2\rangle\Sigma^\gamma(x,Q^2)\right)
\label{F2andqLO}
\ee
analogous but not identical to the convention defined by eq.~(\ref{F2andq}).
Because of the factor $x$ in eq.~(\ref{F2andqLO}) the  moments of the structure function are related
to the moments of the quark densities by
\be       
\int x^{n-1}\frac{1}{x}F_2^\gamma(x,Q^2)dx=q^{\gamma,n}_{\rm NS}(Q^2)+\langle 
e^2\rangle\Sigma^{\gamma,n}(Q^2)
\label{F2momdef}\ee
or $F_2^{\gamma,m}=q^{\gamma,n}_{\rm NS}+\langle e^2\rangle\Sigma^{\gamma,n}$ with $m=n-1$.

We start with a discussion of the LO result for the pointlike $q^{\gamma,n}_{{\rm PL},\,{\rm NS}}(Q^2)$,
which is given by
\be
q^{\gamma,n}_{{\rm PL},\,{\rm NS}}(Q^2)=\frac{4\pi}{\alpha_S(Q^2)}a^n_{\rm NS}(Q^2)
\label{qgnNSdef}\ee
with
\be
\frac{\alpha_S}{4\pi}=\frac{1}{\beta_0\ln{Q^2/\Lambda^2}}
\ee
and $\beta_0=11-2f/3$. The finding of~\cite{GRV} for  $a^n_{\rm NS}$ can be written as
\be
a^n_{\rm NS}(Q^2)=\frac{\alpha}{2\pi\beta_0}k^n_{\rm NS}\frac{1}{1+d^n_{NS}}\big(1-L^{1+d^n_{NS}}(Q^2)\big)
\label{ansdef}\ee
with
\begin{eqnarray}
 k^n_{\rm NS}&=&3f(\langle e^4\rangle-\langle e^2\rangle^2)2h^n_{\rm QM},\label{knsdef}\\
d^n_{NS}=d^n_{qq}&=&\frac{4}{3\beta_0}\left(4\sum_{j=1}^n\frac{1}{j}-3-\frac{2}{n(n+1)}\right)
\label{dqqdef}\end{eqnarray}
and $L(Q^2)=\alpha_S(Q^2)/\alpha_S(Q_0^2)$ giving the ratio of the strong coupling constant
at $Q^2$ and the starting point $Q_0^2$ of the evolution.
In these equations the effect of gluon radiation on the quark model prediction is cast into a
simple form. The quark model term  $k^n_{\rm NS}$ in eq.~(\ref{ansdef})
is multiplied by a factor accounting for 
the quark quark splitting  in the $q\rightarrow qg$ process. The last term in eq.~(\ref{ansdef})
gives a precise meaning to the so called asymptotic solution: $a^n_{\rm NS}$ becomes independent of
$Q^2$ and $Q_0^2$ in the limit $L\rightarrow 0$ i.e. for $Q^2\rightarrow \infty$.

Similar (albeit more complicated) relations hold for $\Sigma^{\gamma,n}_{\rm PL}(Q^2)$:
\be
\Sigma^{\gamma,n}_{\rm PL}(Q^2)=\frac{4\pi}{\alpha_S(Q^2)}a^n_\Sigma(Q^2)=
\frac{4\pi}{\alpha_S(Q^2)}(a^n_+(Q^2)+a^n_-(Q^2))
\label{Signdef} \ee
with
\be
a^n_\pm(Q^2)=\frac{\alpha}{2\pi\beta_0}k^n_q\frac{d^n_{qq}-d^n_\mp}{d^n_\pm-d^n_\mp} 
\frac{1}{1+d^n_\pm}\big( 1-L^{1+d^n_\pm}(Q^2)\big)
\label{asigdef}\ee
and 
\begin{eqnarray}
 k^n_q&=&3f\langle e^2\rangle2h^n_{\rm QM}\label{kqdef}\\
d^n_\pm&=&\frac{1}{2}\left(d^n_{qq}+d^n_{gg}\pm \sqrt{(d^n_{qq}-d^n_{gg})^2-4 d^n_{qg}d^n_{gq}}
\,\,\right)\enspace .
\label{dppdef}\end{eqnarray}
The dependence of  $a^n_{\rm NS}$ and $\Sigma^{\gamma,n}$ on the parameter $Q_0^2$ is not
expressed explicitly on the left hand side of  eq.~(\ref{ansdef}) and eq.~(\ref{asigdef}).
The most important consequence of this dependence is the vanishing of the parton densities 
at the starting scale because $L=1$ at $Q^2=Q_0^2$.

All splitting terms $d^n_{rr'}$ required for the evaluation of equations~(\ref{ansdef})  and~(\ref{asigdef}) can be derived
from the splitting function moments $P^{(0)n}_{rr'}$ in~\cite{GRV} 
(and reference~\cite{FKL} quoted therein) using $d^n_{rr'}=-4P^{(0)n}_{rr'}/\beta_0$. 
Equation~(\ref{dqqdef}) above may serve as a check of the normalization used in this paper.
The asymptotic solution ($L=0$) for $a^n_\Sigma(Q^2)$ can be cast into the compact form~\cite{Pet1}
\be
a^n_{\Sigma,as}=\frac{\alpha}{2\pi\beta_0}k^n_q\frac{1+d^n_{gg}}{1+d^n_{PP}}
\ee
with
\be
d^n_{PP}=d^n_{qq}+d^n_{gg}+d^n_{qq}d^n_{gg}-d^n_{qg}d^n_{gq}\enspace .
\ee

A very useful combination of the results obtained so far is given by
\be
\int x^{n-2}F_{2,{\rm PL}}^\gamma(x,Q^2)dx=\frac{4\pi}{\alpha_S(Q^2)}\sum_i\frac{A^n_i}{1+d^n_i}
\left(1-L^{1+d^n_i}(Q^2)\right)
\label{F2LOPL}\ee
where $i=NS,+,-$ and
\be
A^n_{NS}=\frac{\alpha}{2\pi\beta_0}k^n_{\rm NS}
\ee
\be
A^n_\pm=\frac{\alpha}{2\pi\beta_0}k^n_q\langle e^2\rangle\frac{d^n_{qq}-d^n_\mp}{d^n_\pm-d^n_\mp} 
\enspace .
\ee
Setting  $L=0$ and all  $d^n_{rr'}=0$ 
with $r,r'\in q,g$  the quark model result
\be
F_{2,{\rm QM}}^{\gamma,m}(Q^2)=\frac{3\alpha}{\pi}f\langle e_q^4\rangle h^n_{\rm QM}
\ln{\frac{Q^2}{\Lambda^2}}
\label{F2QM1}\ee
with $m=n-1$ is retained.

The splitting functions $d^n_{rr'}$ (or anomalous 
dimensions as they are often called following the OPE method) are not restricted to integer
$n$ values. For example, the harmonic sum $\sum^n_1(1/j)$ in eq.~(\ref{dqqdef}) can be interpolated 
with the help of the 
Digamma function $\Psi(n)$. There are more complicated harmonic sums contained in the 
other $d^n_{rr'}$ functions and it is even necessary to continue all $n$ dependent functions
in equations~(\ref{ansdef}) and (\ref{asigdef}) into the complex plane in order to invoke the standard method
of inverting the Mellin moments by evaluating the integral
\be
F_2^\gamma(x,Q^2)=\int_{c-\imath\infty}^{c+\imath\infty} F_2^{\gamma}(m,Q^2)x^{1-m}dm
\label{Mellintegr}\ee
where $m$ is now a continuous complex variable and the contour $c$ has to lie on the right hand side
of the rightmost singularity in $F_2^{\gamma}(m,Q^2)$.

In practice, instead of inverting $a^n_{\rm NS}(Q^2)$ and $a^n_\Sigma(Q^2)$ the linear combinations
\begin{eqnarray}
a^m_{\rm val}(Q^2) &=&\frac{1}{\alpha}\left[\frac{\langle e^4\rangle}{\langle e^4\rangle-\langle e^2\rangle^2}
a^n_{\rm NS}(Q^2)\right] \\
a^m_{\rm sea}(Q^2) &=&\frac{1}{\alpha}\left[\langle e^2\rangle a^n_\Sigma(Q^2)-\frac{\langle e^2\rangle^2}
{\langle e^4\rangle-\langle e^2\rangle^2}a^n_{\rm NS}(Q^2)\right]\nonumber
\end{eqnarray}
were chosen because the shape of the corresponding  \emph{valence} and \emph{sea} distributions in $x$-space is
quite different. The pointlike LO solution in $x$-space is then given by
\be
\frac{1}{\alpha} F_{2,{\rm PL}}^\gamma(x,Q^2)=\frac{4\pi}{\alpha_S(Q^2)}
[a_{\rm val}(x,Q^2)+a_{\rm sea}(x,Q^2)]\enspace .
\label{LOv1}\ee
\begin{table}
{\bf a)}\hspace{75mm}{\bf b)} 

 \begin{minipage}{75mm}
 \begin{tabular}{@{}r|r|r|r|r@{}}
& $a_{\rm val}$\,\,\,&$a_{\rm sea}$\,\,\,&$b_{\rm val}$\,\,\,&$b_{\rm sea}$\,\,\,\, \\
\hline
$\delta$&0.7147&-0.70394&0.1046&-1.45262\\
$\beta$&0.1723&0&0.1036&0\\
$c_0$&0.0167& 0.00011&-0.1199&-0.00024\\
$c_1$&-0.0277&-0.00031&-0.1759&-0.00138\\
$c_2$&0.0361&0.00028&1.2996&-0.00300\\
$c_3$&-0.0010&0.00009&-1.6261&0.01201\\
$c_4$&0&0&0.2478&-0.00740\\
\end{tabular}
\end{minipage} \hfill
\begin{minipage}{75mm}
 \begin{tabular}{r|r|r|r|r}
& $a_{\rm val}$\,\,\,&$a_{\rm sea}$\,\,\,&$b_{\rm val}$\,\,\,&$b_{\rm sea}$\,\,\,\, \\
\hline
$\delta$&0.6953&-0.71529&0.0858&-1.72537\\
$\beta$&0.1761&0&0.1080&0\\
$c_0$&0.0328& 0.00033&-0.1535&-0.00061\\
$c_1$&-0.0534&-0.00093&-0.4943&0.00153\\
$c_2$&0.0695&0.00085&2.6800&-0.02831\\
$c_3$&-0.0193&-0.00026&-3.2036&0.05486\\
$c_4$&0&0&0.5097&-0.02808\\
\end{tabular} 
\end{minipage}

\caption{Coefficients needed to calculate the pointlike asymptotic contribution to $F_2^\gamma$
according to equations~(\ref{LOv1}) and~(\ref{HOv1}) for {\bf a)} 3 flavors (left table) and {\bf b)}  4 flavors
(right table). See text for further details.
Each function ($a_{\rm val}(x)$ etc.) is given by
the sum  $ x^\delta (1-x)^\beta\sum_{i=0}^4 c_ix^i$. Only the first 2 columns of each table are
needed for the LO result.}
\label{tabelle1}\end{table}
We focus on the asymptotic solution ($L=0$), where $a_{\rm val}$ and $a_{\rm sea}$ do not depend on 
$Q^2$ and $Q_0^2$.
It was found advantageously to follow the inversion method outlined in~\cite{Duke}, because it quickly
leads to analytical expressions for $F_2^\gamma(x,Q^2)$ which can be further used in fitting the data.
Using the ansatz $x^\delta (1-x)^\beta\sum_{i=0}^4 c_ix^i$  for $a_{\rm val}(x)$ and $a_{\rm sea}(x)$  
the coefficients $\delta,\beta,c_i$ were  determined by fitting the moments of these model functions
to $a^m_{\rm val}$ and $a^m_{\rm sea}$ with $m=n-1$ respectively. 
The method can easiliy be extended to nonasymptotic solutions by repeating this procedure 
for any given pair of $Q^2$ and $Q_0^2$. 

\begin{figure}[h]
\begin{minipage}{72mm}
\includegraphics[width=\linewidth]{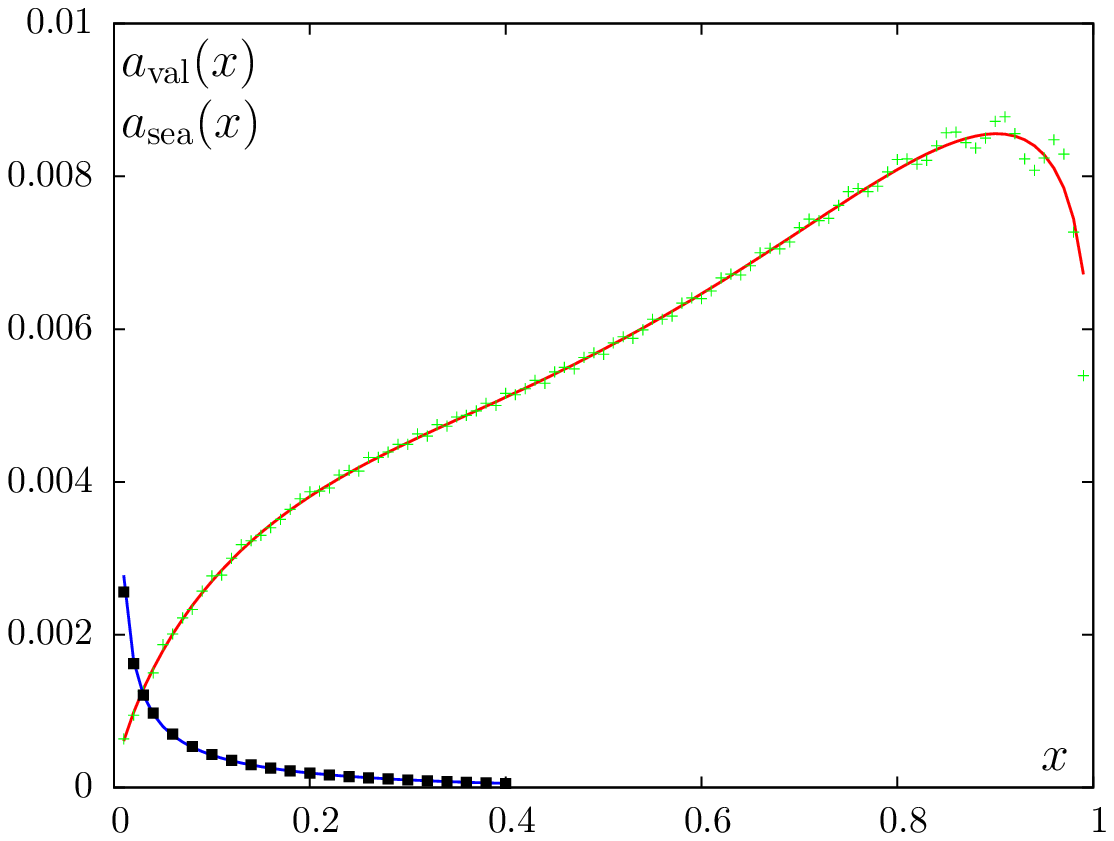}
\caption{Asymptotic pointlike solutions for $f=3$. Upper curve $a_{\rm val}(x)$, lower curve
$a_{\rm sea}(x)$. Both curves are calculated with the fitting method described in the text. In
addition the crosses and boxes represent the numerical evaluation of the integral
(\ref{Mellintegr}) in the complex plane.}
\label{fig:inversion}
\end{minipage}\hfill
\begin{minipage}{72mm}
\includegraphics[width=\linewidth]{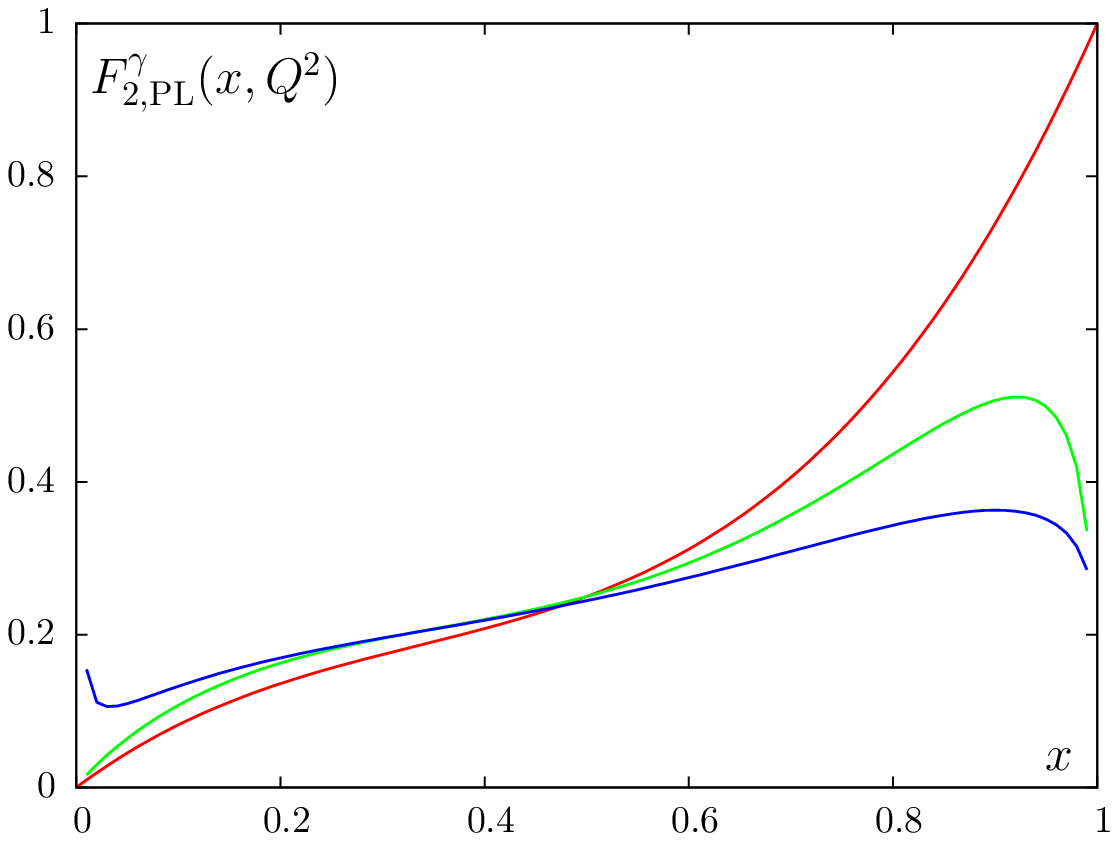}
\caption{Red line: quark model (\ref{qmv1}) in units of
$N_q$ according to eq.~(\ref{Nqdef}). Blue line: Asymptotic  LO QCD prediction in units of
$N_q/\alpha$ using the functions given in table 1. Green line: quark model (in units of $N_q$) 
as defined in eq.~(\ref{qmv2}) including non leading terms in the logarithm. 
 See text for more details.}

\label{fig:modelvgl}
\end{minipage}
\end{figure}

The results of both inversion methods agree very well, which is demonstrated in Fig.~\ref{fig:inversion},
where $a_{\rm val}(x)$ and $a_{\rm sea}(x)$ for $f=3$ and $L=0$  have been plotted.
The lines were obtained using functions modeling 
the moments, whereas the points were calculated  by numerically evaluating the integral (\ref{Mellintegr})
in the complex plane~\cite{Z3}.   

The coefficients necessary for calculating the asymptotic functions 
$a_{\rm val}(x)$ and $a_{\rm sea}(x)$ are listed in table
\ref{tabelle1}a for $f=3$ and table \ref{tabelle1}b for $f=4$. Note that conventional factors 
connecting structure functions and quark densities like in equations~(\ref{F2andq},\ref{F2andqLO}) have been absorbed
in $a_{\rm val}(x)$ and $a_{\rm sea}(x)$.

Obviously the LO QCD calculation (\ref{LOv1}) preserves the $\ln Q^2$ behavior of the quark model but 
changes the $x$ dependence significantly. This is demonstrated in Fig.~\ref{fig:modelvgl} where 
$xh_{\rm QM}(x)$ i.e. the quark model
result (\ref{qmv1}) in units of
\be N_q=\frac{3\alpha}{\pi}\sum_{q=u,d,s}{e_q^4}\ln{\frac{Q^2}{\Lambda^2}}
\label{Nqdef}\ee
is compared to the evaluation  of eq.~(\ref{LOv1}) in units of $N_q/\alpha$.

The fact that the $\ln Q^2$ behavior of the quark model is preserved, is a consequence of the delicate
balance between the increase of the quark population within the photon by the source 
term $\gamma \to \bar{q} q$
and the depletion by gluon radiation $q \to qg$ which however is damped by the decreasing coupling 
due to asymptotic freedom. Would the coupling be fixed at a non-zero value \cite{Pet}, 
then the gluon radiation 
would be so strong that asymptotically the parton densities would fall off to zero as a 
power for any finite value 
$x > 0$. Thus, the $\ln Q^2$ rise of the photon structure function is an 
exciting consequence of asymptotic freedom 
in QCD. For the same reason, the quark population is depleted at large $x$ by gluon radiation, 
and the quarks 
accumulate at small $x$. As a result, the photon structure function is strongly tilted
 -- a  remarkable prediction of perturbative QCD. 

To complete the picture, Fig.~\ref{fig:modelvgl} also contains a QED like variant of the quarkmodel
(\ref{qmv2}) with a log factor $\ln{W^2/m_0^2}$  instead of $\ln{Q^2/\Lambda^2}$. 
With $W^2=Q^2(1-x)/x$ the curve was calculated for $Q^2=100$ GeV$^2$ and $m_0$= $0.3$ GeV and then divided
by $N_q$ using $\Lambda=0.3$. The curve is closer to the QCD prediction depending on the new 
parameter $m_0$.

Finally a careful inspection of the LO QCD result in
Fig.~\ref{fig:modelvgl} reveals a small upward kink beginning at $x\approx 0.02$, because $a_{\rm sea}$ increases $\sim x^{-0.7}$ for 
$x\rightarrow 0$. This divergence can be traced back  to a pole
of $a^n_-$ (\ref{asigdef}) when 
$d^n_-$ approaches  $-1 $  for  $n< 2$. These poles which plague the asymptotic perturbative calculations 
need not concern us as long as they are confined to very small values of $x$. The full solution has no poles because for any finite value of $L$ the quotient $(1-L^{1+d^n_-})/(1+d^n_-)$
in eq.~(\ref{asigdef}) remains finite for $d^n_-\rightarrow -1$.

\subsection{Next to leading order calculations}
In next to leading order the moments of the parton densities are changed, for example
eq.~(\ref{qgnNSdef}) reads now
\be
q^{\gamma,n}_{{\rm pl},{\rm NS}}(Q^2)=\frac{4\pi}{\alpha_S}a^n_{\rm NS}(Q^2)+\tilde{b}^n_{\rm NS}(Q^2)\enspace .
\label{qgnNSdef1}\ee
All NLO effects are contained in $\tilde{b}^n_{\rm NS}(Q^2)$ thus $a^n_{\rm NS}(Q^2)$ is
the leading order result defined in eq.~(\ref{ansdef}). A similar relation holds for $\Sigma^{\gamma,n}(Q^2)$
replacing eq.~(\ref{Signdef}). Besides adding new terms to the parton densities, in NLO the
quark model like relation  eq.~(\ref{F2momdef}) between structure function
and quark densities is also changed. Depending on the factorization
scheme used, products of quark densities and the so called Wilson terms have to be added to the right hand side. The  
lengthy expressions needed to calculate the moments of the structure function 
in the  $\overline{{\rm MS}}$ scheme are again
all contained in~\cite{GRV} and~\cite{FKL}. The results can be nicely cast into the 
form of eq.~(\ref{F2LOPL}) 
\begin{eqnarray}
\int x^{n-2}F_{2,{\rm PL}}^\gamma(x,Q^2)dx&=&\frac{4\pi}{\alpha_S}\sum_i\frac{A^n_i}{1+d^n_i}
\left(1-L^{1+d^n_i}\right)\nonumber\\
&&+\sum_i\frac{B^n_i}{d^n_i}
\left(1-L^{d^n_i}\right)+\sum_i\frac{C^n_i}{1+d^n_i}
\left(1-L^{1+d^n_i}\right)+D^n
\label{F2NLOPL}
\end{eqnarray}
containing  all NLO contributions in the second row. 

For the numerical evaluation  we prefer again to regroup all terms  according to the
\emph{valence} and \emph{sea} scheme. After inverting the moments the  final equation
describing the pointlike solution
\be
\frac{1}{\alpha} F_{2,{\rm PL}}^\gamma(x,Q^2)=\frac{4\pi}{\alpha_S(Q^2)}
[a_{\rm val}(x)+a_{\rm sea}(x)]
+b_{\rm val}(x)+b_{\rm sea}(x)
\label{HOv1}
\ee
is obtained. The strong coupling constant now has to be evaluated in NLO
\be
\frac{\alpha_S}{4\pi}=\frac{1}{\beta_0\ln{Q^2/\Lambda^2}}-
\frac{\beta_1}{\beta_0^3}\frac{\ln\ln{Q^2/\Lambda^2}}{(\ln{Q^2/\Lambda^2})^2}
\ee
with $\beta_1=102-38f/3$. 
For the asymptotic solution the functions 
$a_{\rm val}(x)$, $a_{\rm sea}(x)$, $b_{\rm val}(x)$, $b _{\rm sea}(x)$
can be calculated in a good approximation with the help of
tables~\ref{tabelle1}a,b  for $f=3,4$.
Because $ F_{2,{\rm PL}}^\gamma$ and $\alpha_S$ are defined including non leading terms,
eq.~(\ref{HOv1}) constitutes in the asymptotic regime ($L=0$) an unambiguous QCD prediction  depending on one
parameter ($\Lambda$) only.

Like in the LO case the structure of eq.~(\ref{HOv1}) does not change if non asymptotic solutions are considered. 
One has then, however, for each pair of $Q^2,Q_0^2$-values first to go  through the procedure of calculating and
inverting the moments including now explicitly $Q^2$ dependent factors like in eq.~(\ref{ansdef}). 

\begin{figure}[ht]
\begin{minipage}{72mm}
\includegraphics[width=\linewidth]{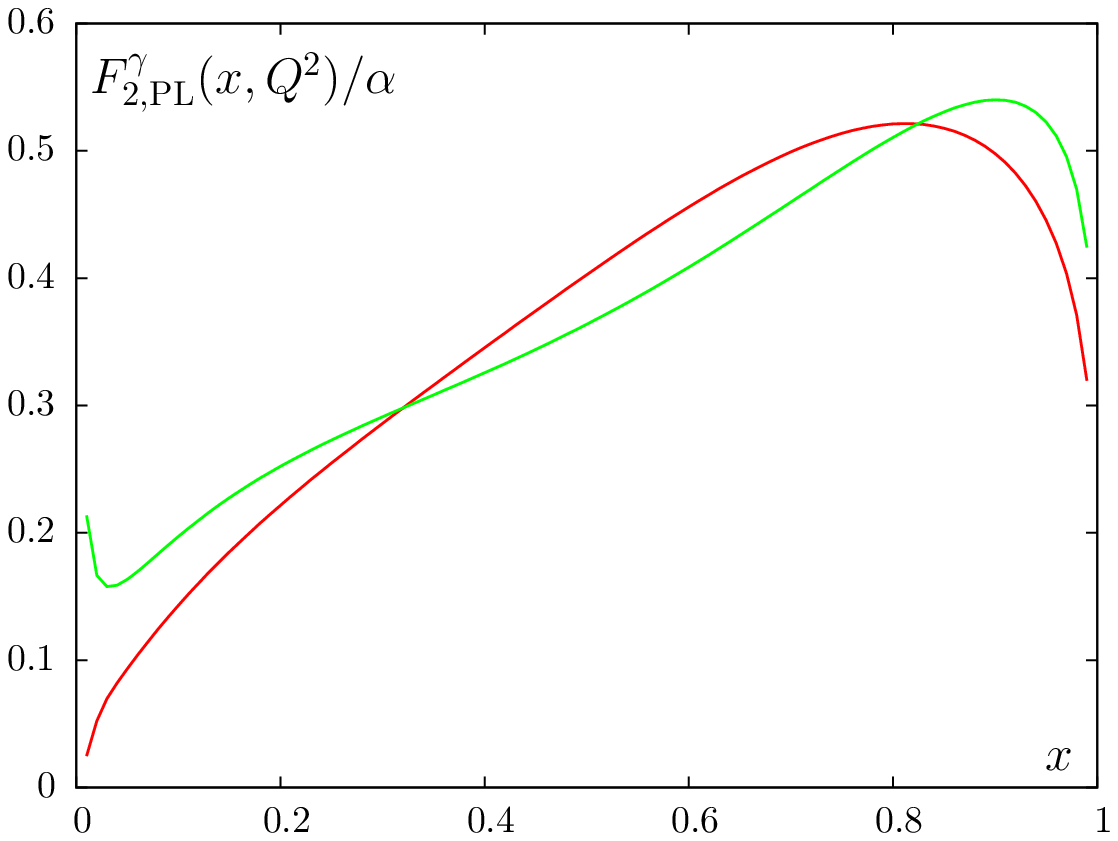}
\caption{Comparison of the asymptotic pointlike structure function 
 in units of $\alpha$ at leading (green curve)
and next to leading order (red curve) QCD for $Q^2=100$ GeV$^2$ and $\Lambda=300$ MeV.} 
\label{fig:NLO-LO}
\end{minipage}\hfill
\begin{minipage}{72mm}

\vspace{5mm}
 \includegraphics[width=\linewidth]{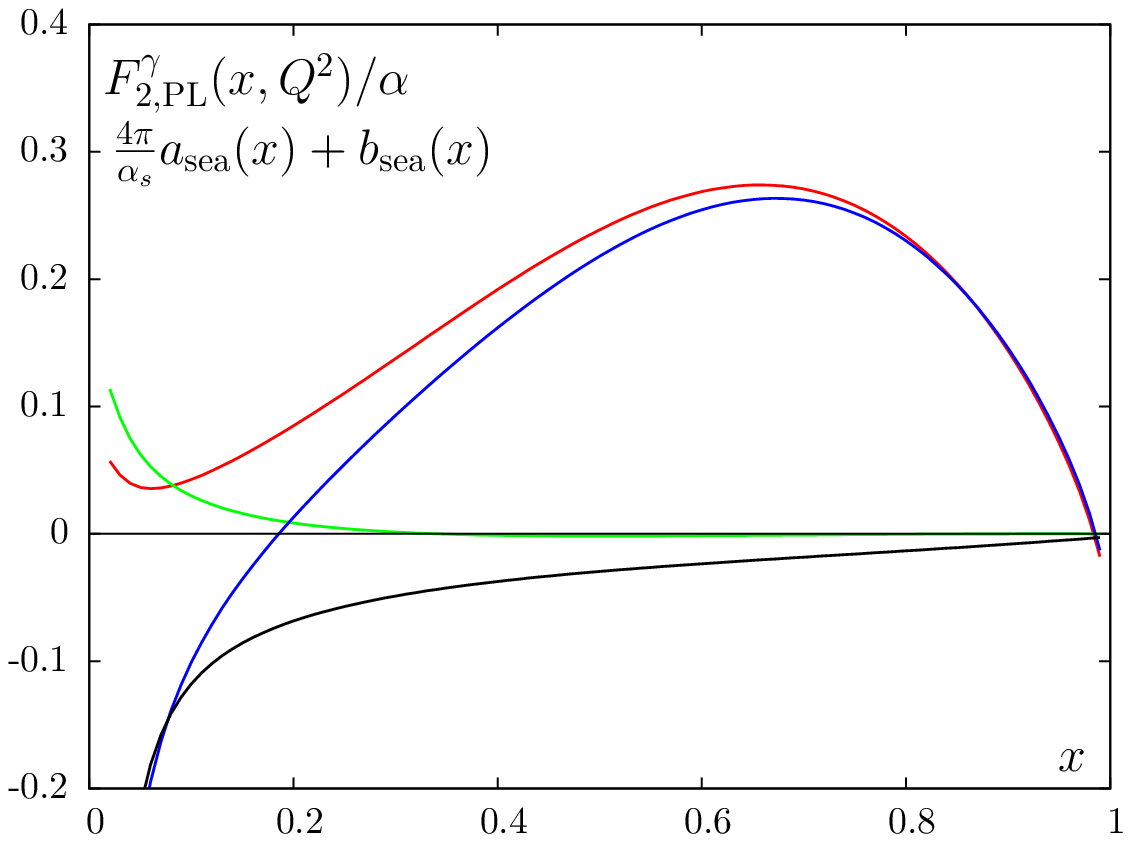}
\caption{Red curve: Asymptotic pontlike structure function
in units of $\alpha$ 
for $f=4$ at $Q^2=3$ GeV$^2$ and $\Lambda=500$ MeV.
Green curve: Sea term $(4\pi/\alpha_S)a_{\rm sea}(x)+b_{\rm sea}(x)$ calculated for the same parameters.
These curves differ qualitatively from the blue and black curve based on incorrect 
moments $b_{\rm sea}^n(Q^2)$. } 
\label{fig:DOvgl}
\end{minipage}
\end{figure}

Due to the negative correction $b_{\rm val}(x)+b_{\rm sea}(x)$ at $x\rightarrow 1$
the region of high $x$ values is further depleted in NLO as can be seen in Fig.~\ref{fig:NLO-LO}
where the asymptotic LO and NLO results are compared for $f=3$ and $Q^2=100$ GeV$^2$
with $\Lambda=300$ MeV. At low $x$ values the NLO correction
is also negative and
would for $x<0.01$ due to the divergence of $b_{\rm sea}(x)$ even lead to a negative unphysical structure function. This time the divergence is caused by $d^n_-=0$ for $n=2$ leading to a pole in $B^n_-/d^n_-$  which for the asymptotic solution is not compensated by the factor $(1-L^{d^n_-})$ 
in~(\ref{F2NLOPL}). The resulting spike is confined to very small $x$-values but is nevertheless
of principal importance because it does not allow the calculation of a sum rule for 
$ F_{2,{\rm PL}}^\gamma(x,Q^2)$ at $L=0$. 

A further example is studied in Fig.~\ref{fig:DOvgl} choosing $f=4$ at $Q^2=3$ GeV$^2$ and $\Lambda=500$
MeV. $ F_{2,{\rm PL}}^\gamma$ is positive in the whole domain considered (red curve) with a positive
sea term $(4\pi/\alpha_S)a_{\rm sea}(x)+b_{\rm sea}(x)$ given by the green curve. 
Due to an unfortunate algebraic error~\cite{BB} the moments $b_{\rm sea}^n(Q^2)$ used in all papers up to 1992
were not correct and resulted in a strongly negative sea term (black curve in Fig.~\ref{fig:DOvgl}) 
which in turn led to a negative pointlike structure function already 
for $x<0.2$~\cite{Duke} i.e.\ much earlier than in the example of Fig.~\ref{fig:NLO-LO}.
It is not surprising that this finding caused a lot of concern in 
the literature.

\subsection{Master formula and the problem of singularities}
As already shown by Witten~\cite{Witt} the moments of the photon structrure function contain besides the pointlike piece an additional term which in lowest order is written as $\sum_i L^{d^n_i}H_i^n(Q_0^2)$
showing the characteristic hadronic $Q^2$ dependence.
The functions $H_i^n(Q_0^2)$ can, however, not be calculated perturbatively. 
Adding the pointlike and hadronic pieces the resulting formula
\begin{eqnarray}
\int x^{n-2}F_2^\gamma(x,Q^2)dx&=&\sum_i H_i^nL^{d^n_i}
+\frac{4\pi}{\alpha_S}\sum_i\frac{A^n_i}{1+d^n_i}\left(1-L^{1+d^n_i}\right)\nonumber\\
&&+\sum_i\frac{B^n_i}{d^n_i}
\left(1-L^{d^n_i}\right)+\sum_i\frac{C^n_i}{1+d^n_i}
\left(1-L^{1+d^n_i}\right)+D^n
\label{master0}\end{eqnarray}
determines the moments of the structure function for  $Q^2> Q_0^2$.
The functions $H_i^n(Q_0^2)$ are either calculated from a fit to a structure function measured  at
some low input scale $Q^2=Q_0^2$ or taken from hadronic models like vector meson dominance (VMD) with an input scale around $0.5$ GeV$^2$ (see next section). 
In any case~(\ref{master0}) is free of singularities but with  $Q_0^2\gg \Lambda^2$ the sensitivity to $\Lambda$ is much reduced. In order to obtain  an
equation containing an absolute prediction which is sensitive to the QCD scale parameter one has to set
$L=0$ in the pointlike part above. Instead of simply doing this by hand we investigate the conditions which allow this procedure. 

The pointlike terms can be rewritten as
\begin{eqnarray}
F_{2,{\rm PL}}^{\gamma,m}(Q^2)&=&\frac{4\pi}{\alpha_S(Q^2)}\left[\sum_i\frac{A^n_i}{1+d^n_i}+\frac{\alpha_S(Q^2)}{4\pi}
\sum_i\frac{C^n_i}{1+d^n_i}\right]+\sum_i\frac{B^n_i}{d^n_i}+D^n\\
&&-\frac{4\pi}{\alpha_S(Q_0^2)}\sum_i  
\left[\frac{A^n_i}{1+d^n_i}+\frac{\alpha_S(Q^2)}{4\pi}\sum_i\frac{C^n_i}{1+d^n_i}\right]L^{d^n_i}(Q^2)
-\sum_i\frac{B^n_i}{d^n_i}L^{d^n_i}(Q^2)\nonumber
\end{eqnarray} 
The terms proportional to $A_i^n$ and $B_i^n$ in the second row showing the typical hadronic $Q^2$ dependence can be combined with the first term in~(\ref{master0}) into a new hadronic contribution
$\sum_i L^{d^n_i}\tilde{H}_i^n$.  The singularities connected with the $C_i$ term are damped by a factor $\alpha_S/4\pi$ but can alo be systematically absorbed into the original higher order (h.o.)
hadronic terms. We thus find
\be
F_{2}^{\gamma,m}(Q^2)=\sum_i \left(\tilde {H}_i^n(Q_0^2)L^{d^n_i}(Q^2)+{\rm h.o.}\right)
+\frac{4\pi}{\alpha_S(Q^2)}\sum_i\frac{A^n_i}{1+d^n_i}
+\sum_i\frac{B^n_i}{d^n_i}+\sum_i\frac{C^n_i}{1+d^n_i}+D^n\enspace .
\label{master1}\ee
In this sum of hadronic terms and the asymptotic solution $F_{2,{\rm PL}}^{\gamma,m}(L=0)$ 
the new hadronic piece 
contains divergencies which exactly cancel the divergencies
of the asymptotic solution~\cite{DuAAc,CBW}. The basic assumption for comparison with data is then to identify the new hadronic piece for large enough $x$-values (say $x>0.1$) with the VMD parameterization of section 3.4, which is certainly only justified if the spikes are confined to very small $x$.

We have shown this assumption to be valid for the LO and the NLO calculations. However in
NNLO  a completely different situation is to be faced. The most dangerous singularities originate now from 
NNLO terms
${\cal D}_n^i=G _i/(1-d^n_i)$ which have to be added on the right hand side of (\ref{master1}). As an example we study the behaviour of 
$d^n_-$ which for $f=3$ approaches 1 for $n=6.0445$. In the vicinity of the pole
at $n=n_0$ we write
\be
{\cal D}^n_-\approx\frac{c}{n-n_0}\enspace 
\ee
which leads in $x$-space to a divergent term $\sim c/x^{5.0445}$. The coefficient $c$ can be estimated from table II of~\cite{USU}. Using ${\cal D}_-^6=-4007$ we get $c\approx 178$ resulting 
after multiplication with $\alpha_S/4\pi$ in 
a contribution $\Delta F_2^\gamma\approx 3/x^5$ to the structure function at small $x$.
This huge singularity is obviously not confined to small $x$-values and makes (together with additional singularities) the prediction of the asymptotic $F_{2,{\rm PL}}^\gamma(x,Q^2)$ unreliable.

The principal problem of the poles  of $1/(1+d^n_i)$, $1/d^n_i$ and $1/(1-d^n _i)$ in the LO, NLO and NNLO
evaluation of  $F_{2,{\rm PL}}^{\gamma,m}(L=0)$ is known since long~\cite{Rossi}. But only after 
the necessary explicit three loop QCD calculations 
had been performed~\cite{MVV1,MVV,VMV} it became clear that the residues of the NNLO-poles are not small enough to avoid a contamination of the large $x$-region. Consequently only
the full pointlike solution $F_{2,{\rm PL}}^\gamma (x,Q^2)$ (starting at $Q_0^2=1$\,\,GeV$^2$) hase been calculated
beyond the next to leading order~\cite{MVV1}.

\subsection{Modelling the hadroncic piece of $ F_2^\gamma$}
\begin{figure}[ht]
\begin{minipage}{72mm}
\includegraphics[width=\linewidth]{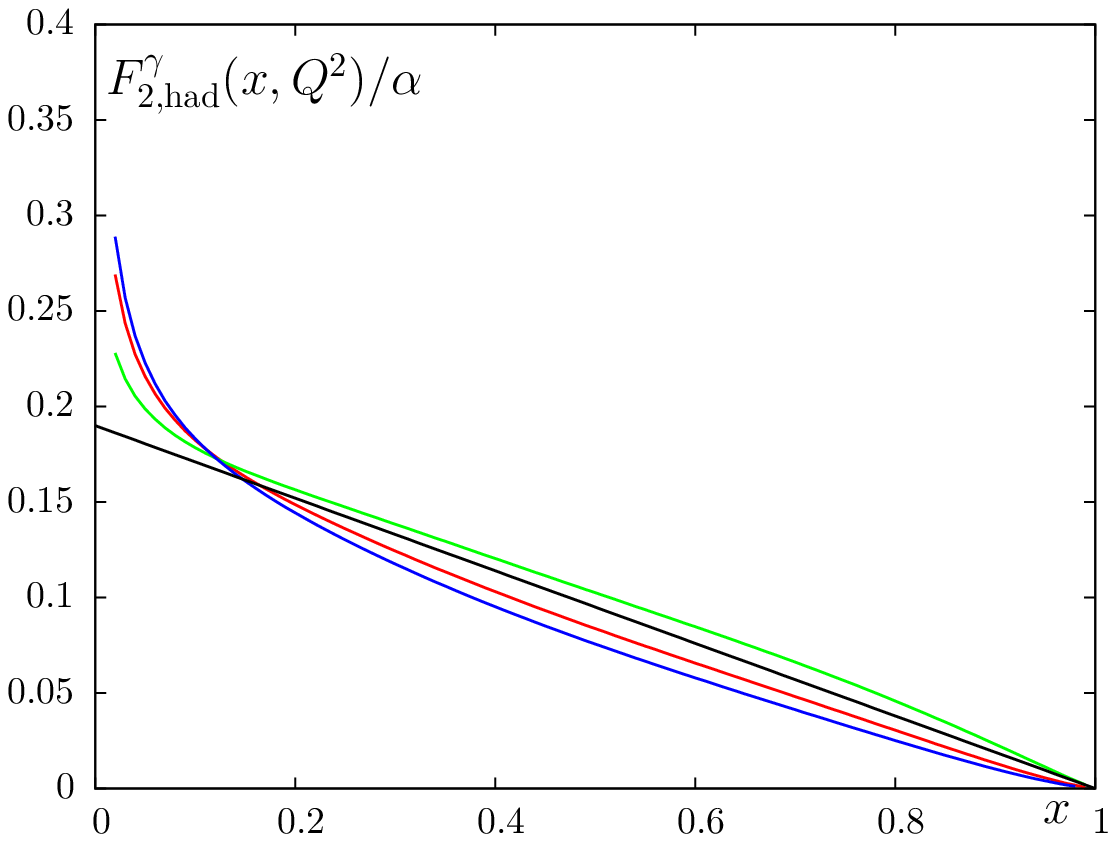}
\caption{Shown is the $x$ dependence of $F^\gamma_{2,{\rm had}}/\alpha$ according to 
~\cite{GRS1} for $Q^2=10,100,400$ GeV$^2$ (green, red and blue curves) 
in comparison with the traditional straight
line (black) ansatz $0.19\,(1-x)$.}
\label{fig:F2had}
\end{minipage}\hfill
\begin{minipage}{72mm}
\vspace{-9mm}
\includegraphics[width=\linewidth]{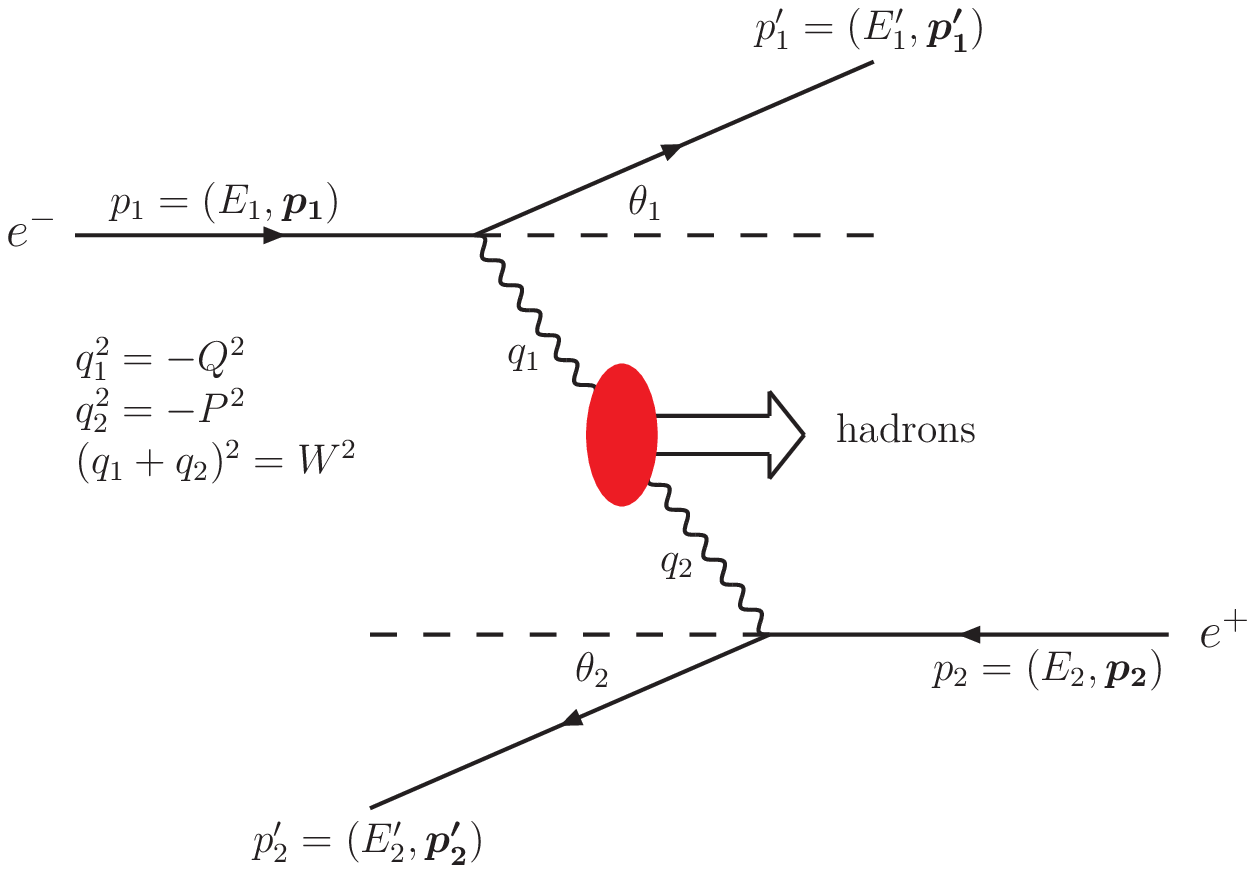}
\caption{Kinematics of the two photon process}
\label{fig:2gamkin}
\end{minipage}
\end{figure}

The coupling of the photon to the final-state hadrons is mediated by quarks and antiquarks. 
If the transverse momentum $k_\perp$ in the splitting process  $\gamma \to q \bar{q}$ is small, 
quark and antiquark travel for a large distance $\tau = \sqrt{s}/k^2_\perp$ 
almost parallel with the same velocity so that strong interactions can develop and bound states
form eventually. Associating a light vector meson with this hadronic quantum 
fluctuation, the corresponding component of the photon wave-function is described by
\begin{equation}
   \ket{\gamma} = \frac{\sqrt{\alpha\pi}}{\gamma_\rho}\sqrt{2} \left(\,  \frac{2}{3} |u \bar{u} \rangle 
                                                   - \frac{1}{3} |d \bar{d} \rangle
                                                   - \frac{1}{3} |s \bar{s} \rangle \, \right) \,,
\end{equation}
which is identical to the vector meson dominance (VMD) ansatz describing the hadronic nature
of the photon

\begin{equation}
   |\gamma \rangle = \frac{\sqrt{\alpha\pi}}{\gamma_\rho}\ket{\rho}+
\frac{\sqrt{\alpha\pi}}{\gamma_\omega}\ket{\omega}+\frac{\sqrt{\alpha\pi}}{\gamma_\phi}\ket{\phi}
\end{equation}
if the photon vector meson couplings $\gamma_V$ are taken from the quark model
neglecting mass effects.
Preferring the measured couplings $\gamma_\rho, \gamma_\omega, \gamma_\phi$
as determined from the partial decay width
$\Gamma^V_{e^+e^-}$ and utilizing isospin invariance~\cite{GRS}
$F^\gamma_{2,{\rm had}}$ has been tied to the well known
pionic quark densities which are available in an easy to use parameterization~\cite{GRS}.

The result for $F^\gamma_{2,{\rm had}}/\alpha$ 
is shown in Fig.~\ref{fig:F2had} for $Q^2=10,100,400$ GeV$^2$. 
Above $x=0.1$ there is little variation with $Q^2$ whereas below $x=0.1$ the sharp increase 
already indicates a tendency to cancel negative spikes in $F^\gamma_{2,{\rm PL}}(x,Q^2)$. It is interesting
to see, how close a simple straight line  $F^\gamma_{2,{\rm had}}/\alpha=0.19(1-x)$ approaches
the results of the complicated evolution model at $x\ge 0.1$. Straight line models of this sort have been used
in the early experimental papers~\cite{Pluto2}.


\section{Two photon kinematics}
Experiments measuring the photon structure function
have until now only been performed at $e^+e^-$ storage rings. The reaction 
\be
e^++e^-\rightarrow e^++e^-+{\rm hadrons}
\ee
is dominated by the so called \emph{two photon diagram} shown in Fig.~\ref{fig:2gamkin}  
which also includes some kinematical definitions. Originally these reactions have  been
considered only as a background to $e^+e^-$ annihilation ($e^+e^-\rightarrow {\rm hadrons}$)
but became, due to the absence of high energetic real photon beams, the only source of
direct experimental information about  $F_2^\gamma(x,Q^2)$.

The incoming leptons in Fig.~\ref{fig:2gamkin}
radiate virtual photons with four momenta $q_1,q_2$ producing a hadronic system $X$ 
with an invariant mass $W=\sqrt{(q_1+q_2)^2}$ The sixfold differential
cross section $d^6\sigma/d\bm{p}_1'd\bm{p}_2'$ is given by a complicated combination of kinematical 
factors and six in principle unknown hadronic functions (four cross sections
and two interference terms) depending on $W^2,Q^2, P^2$. The general formalism
has been discussed in great detail in~\cite{Budnev}, for a recent review and extension see~\cite{Schb}. The 
paper of Budnev et al.~\cite{Budnev} served as the basis of all experimental analyses.

In the limit $P^2\rightarrow 0$ which is realized by very small forward scattering angles of 
one of the incoming leptons (e.g. the positron) the relevant formulae are greatly simplified and
the cross section reads 
\be
\frac{d^2\sigma}{d\Omega_1dE_1'}=\Gamma_t(\sigma_{TT}+\varepsilon\sigma_{LT})f_{\gamma/e}dz
\label{xsecgamt}\ee
with
\be
\Gamma_t=\frac{\alpha}{2\pi^2}\frac{E_1'}{Q^2}\frac{1+(1-y)^2}{y}\\
\ee
\be
\varepsilon(y)=\frac{2(1-y)}{1+(1-y)^2}
\label{epsilon}\ee
and

\be
f_{\gamma/e}(z,\theta_{2,{\rm max}})=\frac{\alpha}{\pi z}\left([1+(1-z)^2]
\label{fgam}
\ln{\left(\frac{E(1-z)}{mz}\theta_{2,{\rm max}}\right)}-1-z\right)
\ee
where the definition $z=(E_2-E_2')/E_2$ has been used.

The two photon cross sections $\sigma_{TT}$ and $\sigma_{LT}$ depend on $Q^2$ and $W^2$. 
The indices represent the transverse ($T$) and longitudinal ($L$ ) polarization of the virtual photons.
The physical interpretation of these equations is like follows: the incoming positron is replaced by a 
beam of quasi real photons with transverse polarization 
traveling along the positron direction. The number of photons in the energy
interval $dz$ is given by $f_{\gamma/e}dz$. The term  $\Gamma_td\Omega_1dE_1'$
denotes the number of 
transversely polarized photons radiated from the electron scattered into the solid angle and energy
interval $d\Omega_1dE_1'$ at angles $\theta_1\gg\theta_2$. With the help of the 
polarization parameter $\varepsilon$ the number of longitudinal photons is given
by $\varepsilon\Gamma_t$.
A very useful feature of this formalism is the factorization of the flux factors into
$\Gamma_t\cdot f_{\gamma/e}$ and $\varepsilon\Gamma_t\cdot f_{\gamma/e}$ where $\Gamma_t$ and
$\varepsilon$
depend on electron variables and $ f_{\gamma/e}$ on positron variables only. This allows
for a considerable simplification  calculating the photon fluxes in Monte Carlo routines
simulating the experiments.
It has been shown~\cite{CBJF} that for $\theta_2<20$ mrad the numerical difference in evaluating
the incoming photon densities from this approach or from the exact formula~\cite{Budnev} is 
less than $0.5$\% for $W/2E>0.05$.

Replacing the cross sections  $\sigma_{TT}, \sigma_{LT}$ by the structure functions
\begin{eqnarray}
&&F_2^\gamma(x,Q^2)=\frac{Q^2}{4\pi\alpha^2}(\sigma_{TT}+\sigma_{LT})\hspace{1cm}\\
&&F_L^\gamma(x,Q^2)=\frac{Q^2}{4\pi\alpha^2}\sigma_{LT}\nonumber
\end{eqnarray}

we arrive after a change of variables at
\be
\frac{d^3\sigma}{dQ^2dxdz}=\frac{2\pi\alpha^2}{xQ^4}\big([1+(1-y)^2]
F_2^\gamma-y^2F_L^\gamma\big)f_{\gamma/e}
\ee
which corresponds to eq.~(\ref{dsigdq2dx}) multiplied by the spectral density of the incoming photons.
Under actual experimental 
conditions, $y$ is quite small in general, so that $F_L^\gamma$ is very difficult to measure. 
Experiments usually focus on the measurement of $F_2^\gamma$ and neglect $F_L^\gamma$.
This is theoretically backed further by the fact 
that quark model and QCD predict $F_2^\gamma$ to be the leading component. 

The standard expression~(\ref{fgam}) has first been derived by Kessler~\cite{Kessler}.
In the spirit of the leading log approximation it can be replaced by
\be
f_{\gamma/e}(z,\theta_{2,{\rm max}})=\frac{\alpha}{\pi z}[1+(1-z)^2]
\ln{\frac{E}{m}}
\label{fgam2}\ee
useful for rough estimates of the counting rate. One has, however, to keep in mind
that neglecting the cutoff 
$\theta_{2,{\rm max}}$ has not only numerical consequences, but quickly violates the basic 
assumption $P^2\approx 0$. 
It is difficult to quote unambiguous limits for the maximum allowed
mean $P^2$ values. Detailed calculation of $\mu$ pair
production~\cite{Nisius} revealed that for $E=45$ GeV and 
$\theta_{2,{\rm max}}=27$ mrad one has $\langle P^2\rangle=0.04$ GeV$^2$ and
$97\%$ of the cross section
is contained in eq.~(\ref{xsecgamt}), a result which is likely also to be valid
for the quark model and QCD. 
It follows that the experiments need 
forward spectrometers with electromagnetic calorimeters very close to the beam pipe
which allow to reject positrons with angles larger than about $25$ mrad via the method
of antitagging.

The basic experimental procedure is thus given by investigating the reaction
$e^+e^-\rightarrow e^+e^-+{\rm hadrons}$ with the electron scattered at angles
larger than $\theta_{2,{\rm max}}$ and the positron traveling undetected down the beam pipe (and vice versa).
Due to the unknown energy $E_2'$ of the outgoing positron $E_\gamma$ is also not known and
the usual relation $Q^2=xys$ cannot be used for calculating $x$. Therefore hadronic 
calorimeters reaching down 
to small forward scattering angles are needed in order to measure the invariant mass $W$ 
of the produced hadronic system and calculate $x$ from eq.~(\ref{xydef}). Unfortunately
the remnants of the
antiquark in Fig.~\ref{fig:Feyn} will also be dominantly concentrated at small angles and losses in the 
hadronic energy are unavoidable. With $W_{\rm exp}\leq W_{\rm true}$ sophisticated unfolding
methods have to be employed in order to reconstruct $x$. These methods are described in 
the experimental publications and reviewed in~\cite{CBW} and~\cite{Nisius}. A discussion 
of the various Monte Carlo routines used by the different experimental groups in evaluating 
the cross section can e.g. be found in~\cite{Nisius}.

\section{Experimental analysis}
Following the pioneering work of the PLUTO collaboration~\cite{Pluto1} many experiments have 
been performed at all high energy $e^+e^-$ storage rings.
In order to avoid the region of small $x$ with its mixture of correlated
hadronic and pointlike contributions we exclude data with $x\leq 0.1$.
Data where the charm component has been subtracted and all data  published in conference proceedings only
are also discarded. Only the most recent publication of statistically overlapping data of the same collaboration was accepted. 
This selection leads to 109 experimental values of $F_2^\gamma(x,Q^2)/\alpha$
with $Q^2$ values ranging from $4.3$ GeV$^2$ to $780$ GeV$^2$
from the 
collaborations ALEPH~\cite{ALEPH1,ALEPH2}, AMY~\cite{AMY1,AMY2}, DELPHI~\cite{DELPHI}, JADE~\cite{JADE},
L3~\cite{L31,L32,L33,L34}, OPAL~\cite{OPAL1,OPAL2,OPAL3}, PLUTO~\cite{Pluto2,Pluto3}, TASSO~\cite{TASSO},
TOPAZ~\cite{TOPAZ} and TPC/2$\gamma$~\cite{TPC}. In cases where the experimental uncertainties could only
be read off the figures the tables of Nisius~\cite{Nisius} were used.
As usual the experimental $x$ and $Q^2$ values are obtained from an averaging procedure over the sometimes
rather large $x$ and $Q^2$ bins. In most cases $x$ coincides with the bin center. 

After the 1980 crisis of the perturbative calculation most QCD analyses were performed like
in deep inelastic scattering by comparing the data to models obtained by evolving the parton
densities from a starting scale $Q_0^2\gg\Lambda^2$ up to $Q^2$. For a recent extensive study
see~\cite{CJK}. On the other hands side data at high $Q^2$ and high $x$ (defined
by $Q^2\ge 59$ GeV$^2$ and $x\ge 0.45$) were fitted to the asymptotic pointlike solution 
(\ref{HOv1}) for $f=3$, supplemented by 
the quark model formula (\ref{F2QED}) for a charm quark
with mass $1.5$ GeV~\cite{AKS}. The fit described the data very well and
resulted in $\alpha_S(M_Z)=0.1183\pm 0.0058$ with $\chi^2=9.1$ for 20 experimental data.

Here we follow a more radical approach and fit the whole sample of 109 data sets to a model
whose three components have been discussed in the previous sections:

\begin{enumerate}
 \item The pointlike asymptotic NLO QCD  prediction  
for 3 light flavors in the $\overline{{\rm MS}}$ scheme with $\Lambda_3$ as the only free 
parameter using the coefficients of table~\ref{tabelle1}a.
\item A quark model calculation of the charm and bottom quark contribution using eq.~(\ref{F2QED}) multiplied
by $3e_q^4$.  
Applying the  $\overline{{\rm MS}}$ scheme for the light quark QCD calculation it is only
natural to use the  $\overline{{\rm MS}}$ masses $M_c=1.275\pm 0.025$ GeV
and $M_b=4.18\pm 0.03$ GeV as quoted by the Particle Data Group~\cite{PDG}.
\item A detailed parameterization (VMD)
for the hadronic part of the structure function~\cite{GRS} including the $Q^2$ evolution.
Examples for different values of $Q^2$ are displayed in Fig.~\ref{fig:F2had}.
\end{enumerate}


Fitting the data with this model results in a value of $\Lambda_3=0.338\pm 0.020$ GeV. The
quality of the fit, measured in terms of the $\chi^2$ value per degree of freedom, is given by
$\chi^2_{\rm dof}=78/108$. A value slightly below unity is probably explained by the neglect
of bin to bin correlations in the fitting procedure. These correlations were only given
in six of the seventeen experimental publications used. 

Following the method explained in~\cite{Marciano} $\Lambda_3$
is converted to $\Lambda_5=0.201\pm 0.015$ GeV and therefore 
$\alpha_S(M_Z^2)=0.1159\pm 0.0013$ is obtained in NLO where the error only reflects the
experimental uncertainties. In order to estimate the theoretical error we 
first neglected the bottom quark contribution which changes 
$\alpha_S(M_Z^2)$ by a very small amount (0.0002). Varying $M_c$ within the quoted error of $\pm 0.0025$ GeV resulted in
$\Delta\Lambda_3=\pm 0.007$. The most important source of theoretical uncertainty is the treatment
of the hadronic contribution. This error is hard to estimate. Possible interferences between hadronic and pointlike part
are very likely restricted  to the region $x\le 0.1$ which is excluded by our data selection.
The authors of~\cite{GRS1} emphasize the very good agreement of their result with pionic and photonic structure function data.
Assuming a $10\%$ normalization error for  $F^\gamma_{2,{\rm had}}/\alpha$ yields $\Delta\Lambda_3=\pm 0.042$.
Adding all errors in quadrature the final result is
\be
\alpha_S(M_Z^2)=0.1159\pm 0.0030
\ee
which agrees nicely with the DIS average
$\alpha_S(M_Z^2)=0.1151\pm 0.0022$ and
the world average $\alpha_S(M_Z^2)=0.1184\pm 0.0007$ as given
in~\cite{PDG,Bethke}.

\begin{figure}
\begin{center}
\includegraphics[width=0.7\textwidth]{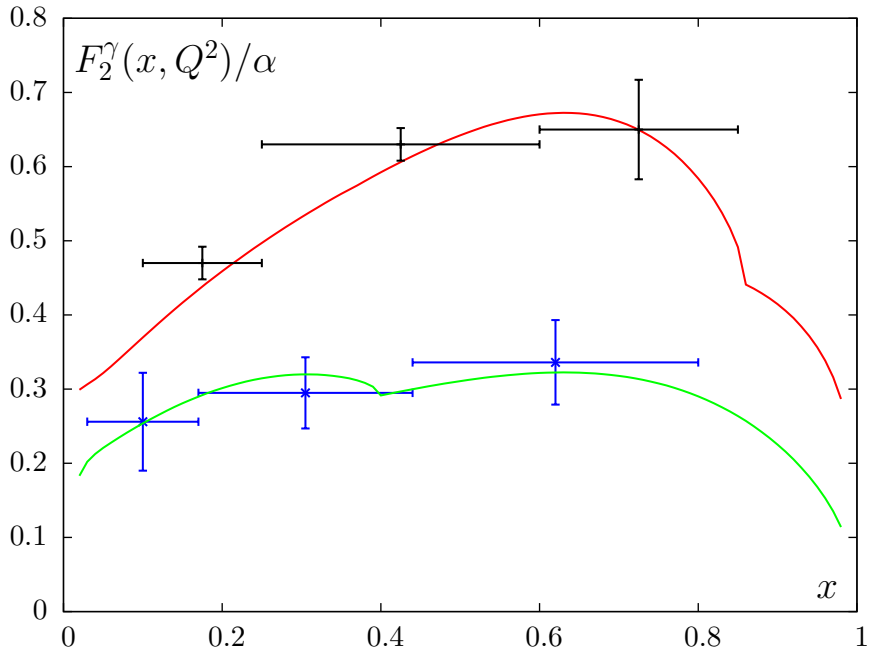}
\end{center}
\caption{Dependence of $F_2^\gamma/\alpha$ on $x$ for two different
values of $Q^2$. The PLUTO data~\cite{Pluto2} at $4.3$ GeV$^2$ (blue crosses) and the OPAL 
data~\cite{OPAL3} at $39.7$ GeV$^2$ (black crosses) are compared with the QCD model described in the text
(green and red curves).}
\label{fig:xplot}
\end{figure}
In order to visualize the impressive agreement between data and theory two examples are 
presented. In Fig.~\ref{fig:xplot} the PLUTO data~\cite{Pluto2} at $4.3$ GeV$^2$ (black crosses)
and the OPAL data~\cite{OPAL3} at $39.7$ GeV$^2$ (blue crosses)
 are compared with our model.  The data clearly do not follow 
the typical mesonic $1-x$ dependence and also demonstrate implicitly the rather strong
$Q^2$ dependence predicted by QCD. 

\begin{figure}
\begin{center}
\includegraphics[width=0.7\textwidth]{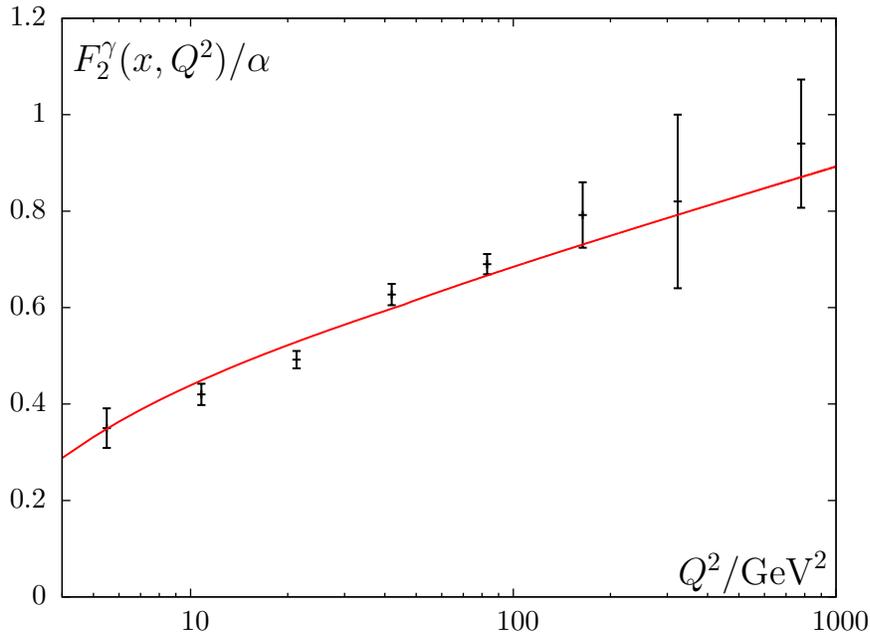}
\end{center}
\caption{Dependence of $F_2^\gamma/\alpha$ on $Q^2$. All available data for $F_2^\gamma/\alpha$ with
$0.3 < \bar{x} <0.5 $ are averaged in
$Q^2$ bins with equal width in $\ln Q^2$ as explained in the text. The red curve shows the result of our QCD model.}
\label{fig:Q2plot}
\end{figure}

Next the $Q^2$ dependence of  $F_2^\gamma(x,Q^2$) is directly tested by selecting 
data with $0.3< \bar{x}< 0.5$. The average 
$x$ value $\bar{x}$ was determined taking the mean value of the $x$ intervals
quoted in the experimental papers. The $36$ data sets are grouped in bins
with equal bin width in $\ln Q^2$.  
Each $ F_2^\gamma$ value  is then shifted to the center of the corresponding bin
using the theoretical model and the weighted average of all data within the bin is calculated. 
The result is shown in Fig.~\ref{fig:Q2plot} together with the theoretical curve
calculated for $x=0.4$. 
The increase of the data with  $\ln Q^2$ is clearly seen, especially in contrast
to the well known slight \emph{decrease}
of the proton 
structure function for $x=0.4$ between $Q^2=5$ and
$800$ GeV$^2$.
Neglecting the small $Q^2$ dependence of the hadronic piece the theoretical model can 
in  LO be written as
$F_2^\gamma(x,Q^2)=a(x)+b(x)\ln({Q^2/{\rm GeV}^2})$.  A fit of the data in Fig.~\ref{fig:Q2plot}
 according to this ansatz yields $b(0.4)=0.133\pm 0.008$ thus establishing 
numerically the predicted increase with
$\ln{Q^2}$ beyond any doubt. This value agrees very well with the earlier analysis
of~\cite{Nisius}.
 
\section{Virtual photon structure}
The perturbative calculations have been extended to the region $\Lambda^2\ll P^2\ll Q^2$~\cite{UW1,UW2} 
which is experimentally accessible requesting also the positron being
scattered into finite angles (double tagging). Regarding the theory   one has in the formalism
of~\cite{GRV} simply to replace
the parameter  $Q_0^2$ by the variable $P^2$ for the 
calculation of $F_{2,{\rm PL}}^{\gamma,n}(Q^2,P^2)$ in LO. Ueamtsu and Walsh~\cite{UW2} emphasized
that for virtual photons also the hadronic piece is perturbatively calculable. The required  additional terms are given in~\cite{UW2,USU}. 

\begin{figure}
\begin{minipage}{72mm}
\includegraphics[width=\linewidth]{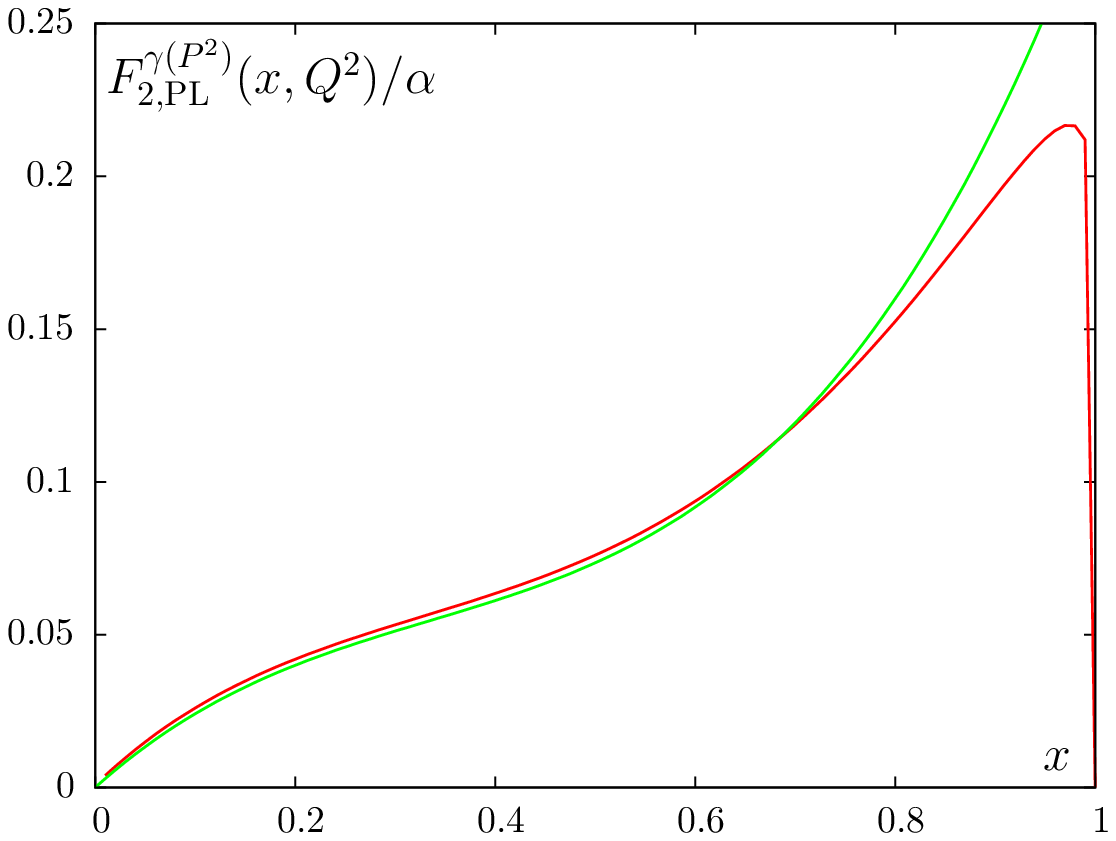}
\caption{Dependence on $x$  of the virtual photon
structure function $F_2^{\gamma(P^2)}(x,Q^2)/\alpha$ calculated in LO for $Q^2=30$ GeV$^2$, $P^2=7.5$ GeV$^2$
 and $\Lambda_3=0.338$ GeV 
(red curve) in comparison (green curve) with the modified quark model result derived from eq.~(\ref{QMP2}).}
\label{fig:F2virtplot}
\end{minipage}\hfill
\begin{minipage}{72mm}
\includegraphics[width=\linewidth]{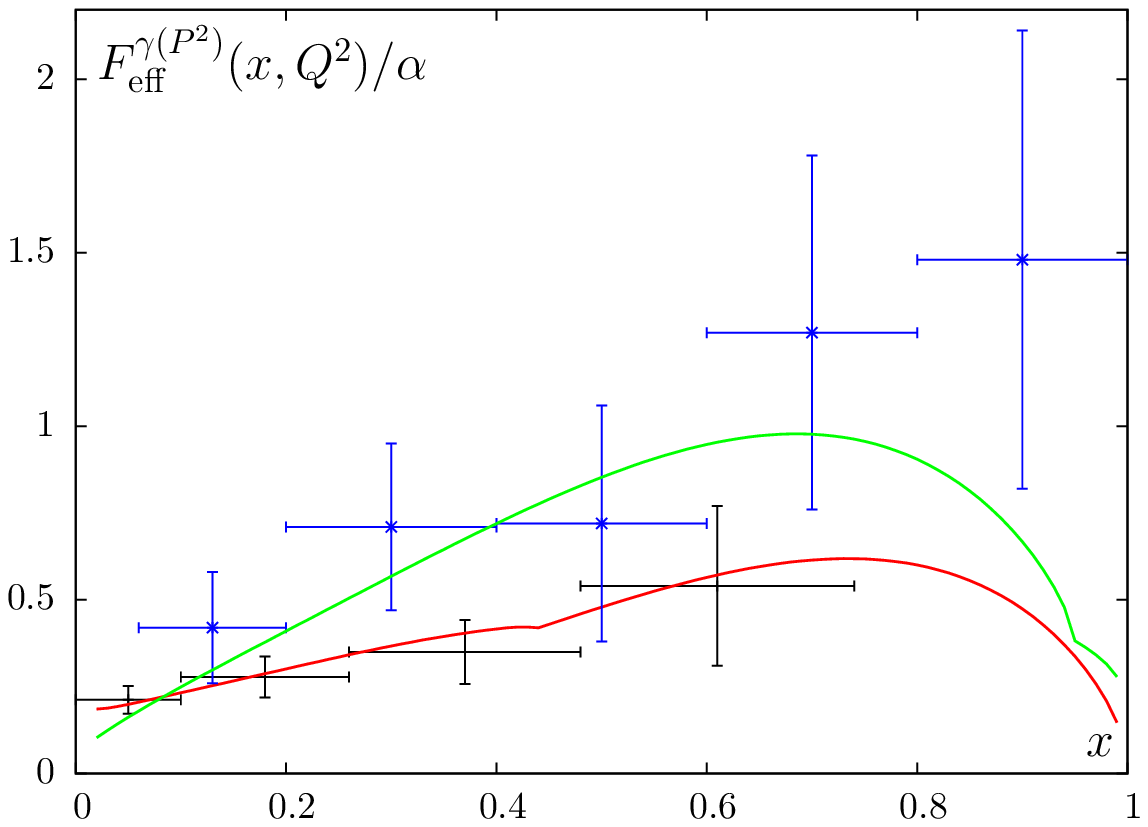}
\caption{Shown is the $x$ dependence of $F_{{\rm eff}}^{\gamma(P^2)}(x, Q^2)/\alpha$. The PLUTO data at 
$Q^2=5$ GeV$^2$, $P^2=0.35$ GeV$^2$  (black crosses)
are compared to the QCD prediction  (red curve) including hadronic and charm quark contributions.
The details are explained in the text. The L3 data at $Q^2=120$ GeV$^2$, $P^2=3.7$ GeV$^2$ (blue crosses)
are compared to the same model (green curve).  
}
\label{fig:F2PLL3virtplot}
\end{minipage}
\end{figure}

Gluon radiation is efficiently suppressed for virtual photons, thus moving $F_2^{\gamma (P^2)}(x,Q^2)$
closer to the quark model
result. Analytically this can be proven easily~\cite{UW1} by investigating the LO order solution, e.g.  $q^{\gamma,n}_{\rm NS}$,
eq.~(\ref{qgnNSdef}). Because in LO $\alpha_S$ is proportional to $1/\ln{(Q^2/\Lambda^2)}$ we have
\be
q^{\gamma,n}_{\rm NS}\sim\frac{\ln{\frac{Q^2}{\Lambda^2}}-\ln{\frac{P^2}{\Lambda^2}}
\left(\frac{\ln{(P^2/\Lambda^2)}}{\ln{(Q^2/\Lambda^2})}\right)^{d_{qq}^n}}{1+d_{qq}^n}\, .
\ee
Using $\ln{(Q^2/\Lambda^2)}=\ln{(P^2/\Lambda^2)}+\ln{(Q^2/P^2)}$ the  nominator reduces in the limit
$\ln{(Q^2/P^2)}\ll \ln{(P^2/\Lambda^2)}$ to $(1+d_{qq}^n)\ln{(Q^2/P^2)}$. Since a similar relation holds for the
singlet term, the final result\footnote{This formula also demonstrates drastically how the introduction of a second scale
destroys the sensitivity to $\Lambda$. In case of the starting  scale $Q^2_0$ of section III with values of
$0.3$ to $1.0$ GeV$^2$ a reduced sensitivity is maintained.} is
\be
F_{2,{\rm PL}}^{\gamma,n}(Q^2,P^2)=\frac{3\alpha}{\pi}f\langle e_q^4\rangle h^n_{\rm QM}
\ln{\frac{Q^2}{P^2}}\enspace ,
\label{QMP2}\ee
i.e. the quark model formula (\ref{F2QM1}) with the log factor replaced by 
$\ln(Q^2/P^2)$.
As an illustration the LO prediction
for $Q^2=30$ GeV$^2$, $P^2=7.5$ GeV$^2$ and $\Lambda_3=0.338$ GeV is compared in Fig.~\ref{fig:F2virtplot}
with eq.~(\ref{QMP2}) showing perfect agreement
at small and medium $x$ values. 
The parameter free QCD prediction (\ref{QMP2}) is, however, difficult to be tested experimentally because with 
the present value of $\Lambda$ the condition 
$\ln{(Q^2/P^2)}\ll \ln{(P^2/\Lambda^2)}$ can hardly be achieved.  

Regarding the determination of $F_2^{\gamma(P^2)}(x,Q^2)$ from the measured
cross section, it should be remembered that the 
virtual photon photon scattering is in general described by four cross sections and two
interference terms (see section IV). 
After proper integration over the interference terms the cross section formula of~\cite{Budnev} can be written as
\be
d\sigma\sim L_{TT}(\sigma_{TT}+\varepsilon_1\sigma_{LT}
+\varepsilon_2\sigma_{TL}+\varepsilon_1\varepsilon_2\sigma_{LL})
\label{xsecgamqq}\ee
where the factor $L_{TT}$ is approximately interpreted as describing the fluxes of transverse virtual photons from the
incoming electron and positron. The polarization parameters $\varepsilon_1$ and 
$\varepsilon_2$ are for typical experimental conditions  close to $1$ 
and therefore 
\be
d\sigma\sim L_{TT}(\sigma_{TT}+\sigma_{LT}+\sigma_{TL}+\sigma_{LL})\enspace .
\ee
The relation between structure functions and cross sections is more complicated than discussed above for 
electron scattering off real photons. In the limit $Q^2\gg P^2$ 
the general formulae given in~\cite{CBW} reduce to
\begin{eqnarray}
F_2^{\gamma(P^2)}(x,Q^2)& =&\frac{Q^2}{4\pi\alpha^2}(\sigma_{TT}+\frac{1}{2} \sigma_{LT})\\
F_L^{\gamma(P^2)}(x,Q^2)&=&\frac{Q^2}{4\pi\alpha^2}\sigma_{LT}\nonumber
\end{eqnarray}
using  $\sigma_{LT}\approx \sigma_{TL}$ and  $\sigma_{LL}\approx 0$.
Defining an effective structure function via
\be
F_{\rm eff}^{\gamma(P^2)}=\frac{Q^2}{4\pi\alpha^2}(\sigma_{TT}+\sigma_{LT}+\sigma_{TL}+\sigma_{LL})
\ee
one gets finally
\be
F_{\rm eff}^{\gamma(P^2)}(x,Q^2)\approx F_2^{\gamma(P^2)}(x,Q^2)+\frac{3}{2}F_L^{\gamma(P^2)}(x,Q^2)\enspace .
\label{F2virtfin}\ee

Experimental data is scarce. The first results of the PLUTO collaboration~\cite{Pluto4} at 
$Q^2=5$ GeV$^2$ and  $P^2=0.35$ GeV$^2$ (black crosses in Fig.~\ref{fig:F2PLL3virtplot})  
were only followed by data of the L3 collaboration~\cite{L33} at
$Q^2=120$ GeV$^2$ and $P^2=3.7$ GeV$^2$ (blue crosses in Fig.~\ref{fig:F2PLL3virtplot}). 

For comparison
with theory $F_2^{\gamma(P^2)}(x,Q^2)$ is calculated in NLO
for 3 flavors choosing $\Lambda_3=0.338$ GeV.
According to eq.~(\ref{F2virtfin})  $F_L^{\gamma(P^2)}(x,Q^2)$ cannot be neglected. The longitudinal structure 
function of real photons is extensively discussed in the literature~\cite{Laenen}.
For virtual photons $F_L^{\gamma(P^2)}(x,Q^2)$ has been calculated in LO and NLO~\cite{UW2,USU}. 
Here we combine the LO result with the NLO calculation of the pointlike and hadronic terms. 
The charm quark contribution for $F_2$ and $F_L$ is taken from the quark model result for real photons as recommended in~\cite{GRS1}. Because the PLUTO data are taken at a $P^2$ value close to the real photon case
a VMD part was added multiplied by 
a form factor $1/(1+P^2/m_\varrho^2)^2$ where $m_\varrho$ is the $\varrho$ meson mass.
Using this form factor the VMD term is 
reduced by a factor 2.5 and thus for the sake of simplicity the straight line model $0.19(1-x)/2.5$
is applied improving somehow the agreement with the data at low $x$.
Altogether the red curve (PLUTO) and the green curve (L3) in Fig.~\ref{fig:F2PLL3virtplot} are in very 
good agreement with the data although admittedly this is not a very decisive test due to the large
experimental errors. A comparison  including the rather small  NNLO corrections can be found in~\cite{SUKU}.

\section*{Conclusions}
Measurements of the photon structure function $F_2^\gamma$ taken at $e^+e^-$ colliders were confronted with 
theoretical models. For real photons the main component is the fixed flavor ($f=3$) NLO asymptotic QCD result in the
$\overline{\rm MS}$ scheme as given in eq.~(\ref{HOv1}) evaluated with the functions of~table \ref{tabelle1}.
This
part is complemented by charm and bottom heavy quark  contributions calculated in the quark model and
by a hadronic contribution taken from vector meson dominance. The model describes not only the $x$ and
$Q^2$ distributions very well but also allows for a precise determination of the strong
coupling constant, yielding $\alpha_S(M_Z^2)=0.1159\pm 0.0030$. 
As explained above the treatment of the hadronic and
the heavy quark contribution to $F_2^\gamma$ does not follow from first principles but is based on  model assumptions. The validity of these assumptions  is supported by the observation that using the standard model value of
$\Lambda_3$ the selected data are described by the model of section 5 with
$\chi^2_{\rm dof}=78.5/108$. Finally the few available data for virtual photons agree well with 
the QCD predictions. 

New experimental input can only be expected from a new high energy $e^+e^-$ collider. At the planned 
linear collider ILC~\cite{ILC} it is in principle possible to install a high intensity beam of real photons via backscattering 
of laser light. This would for the first time allow to study inelastic electron photon scattering 
in a beam of real photons with a spectrum and intensity far superior to the virtual photons used
until now~\cite{ZM}. In addition nagging doubts about the $P^2$ cutoff in some of the two photon 
experiments are baseless in such an environment. 

\small{
\section*{Acknowledgement}
First of all I want to thank  P.M. Zerwas for his constant support and the many 
discussions concerning the theoretical basis. I am  very grateful
for the help I got from R. Nisius. Useful conversations with M. Klasen are also
gratefully acknowledged.

}

\end{document}